\documentclass[aps,prd,reprint,onecolumn,notitlepage,11pt,tightenlines,nofootinbib,showkeys]{revtex4-2}
\pdfoutput=1

\usepackage{lmodern}
\usepackage{microtype}
\usepackage{placeins}

\usepackage[T1]{fontenc}
\usepackage{amsmath,amssymb}
\usepackage{graphicx}
\usepackage{url}
\usepackage{xcolor}
\usepackage[hypertexnames=false]{hyperref}
\definecolor{preprintblue}{RGB}{0,51,102}
\hypersetup{
  colorlinks=true,
  linkcolor=preprintblue,
  citecolor=preprintblue,
  urlcolor=preprintblue,
  pdftitle={Neutrino Mass in Left-Right Symmetry: From Majorana Signals to Dirac Predictions},
  pdfauthor={Vladimir Tello}
}
\newcommand{\refcite}[1]{\cite{#1}}
\newcommand{\Pcal}{\mathcal P}

\newcommand{\eps}{\epsilon}

\newcommand{\dCP}{\delta_{\rm CP}^{\ell}}

\newcommand{\diag}{\mathrm{diag}}
\newcommand{\Tr}{\mathrm{Tr}}

\newcommand{\CP}{\mathrm{CP}}
\newcommand{\CPell}{\mathrm{CP}_\ell}

\begin{document}

\title{Neutrino Mass in Left--Right Symmetry:\texorpdfstring{\\}{ }
From Majorana Signals to Dirac Predictions}

\author{Vladimir Tello}
\affiliation{University of Split, FESB, Croatia}
\date{October 22, 2026}

\begin{abstract}
We review neutrino mass as a probe of restored parity in left--right
symmetric theories, emphasizing the distinct Majorana and Dirac
realizations. In the Majorana theory, lepton-number violation connects
the seesaw mechanism to neutrinoless double beta decay, heavy-neutrino
production and the Keung--Senjanovi\'c process. The Dirac alternative
removes these characteristic signals while retaining a relation between
left- and right-handed flavor. The right-handed leptonic mixing matrix
can then be reconstructed from low-energy data and the spontaneous
parity-breaking parameter. Requiring the two lepton Yukawa matrices to
admit a common real parity basis turns this reconstruction into a
compatibility condition that determines the allowed Dirac phase on each
branch. For three generations, this requirement reduces to a single
rephasing-invariant CP-odd obstruction, which also governs the
right--left modulus mismatch and the relative CP orientation.
The small Dirac masses also restrict the parity deformation and its
induced strong-CP phase. For compressed mixed-sign neutrino spectra, a tiny parity
deformation can nevertheless generate large leptonic CP violation,
despite stringent strong-CP constraints. The resulting branch-dependent
correlations connect leptonic CP, the absolute neutrino mass and strong
CP. We place this construction alongside radiative and unified
approaches to predicting leptonic CP violation.
\end{abstract}

\maketitle

\section{Why neutrino mass belongs to parity}
\label{sec:parity_motivation}

Neutrino mass is often introduced as a small addition to the Standard
Model. From the viewpoint of left--right symmetry, the logical order is
reversed. The starting question is why the weak interaction distinguishes
left from right so completely, and whether this maximal asymmetry may be
the low-energy remnant of a parity-symmetric theory.

The left--right framework implements this idea through an extended
gauge symmetry~\cite{Pati:1974yy,Mohapatra:1974gc,Mohapatra:1974hk,Senjanovic:1975rk}.
We consider the minimal gauge group
\begin{equation}
G_{\rm LR}
=
SU(3)_c
\times SU(2)_L
\times SU(2)_R
\times U(1)_{B-L},
\label{eq:GLR_review}
\end{equation}
with electric charge given by
\begin{equation}
Q
=
T_{3L}
+
T_{3R}
+
\frac{B-L}{2}.
\label{eq:charge_review}
\end{equation}
Quarks and leptons then occur in left- and right-handed doublets,
\begin{equation}
Q_{L,R}
=
(u,d)_{L,R},
\qquad
L_{L,R}
=
(\nu,e)_{L,R},
\end{equation}
which are interchanged by generalized parity. The right-handed neutrino is
required as the parity partner of the observed neutrino, independently of
how its mass is generated. The
chiral weak theory emerges only after \(SU(2)_R\) and parity are
spontaneously broken.

Parity alone neither determines the neutrino-mass scale nor decides whether
neutrinos are Dirac or Majorana particles. It does, however, place that
choice inside the symmetry-breaking problem. If interactions violating
lepton number by two units are present, most notably through scalar
triplets, the right-handed neutrino acquires a Majorana mass and the seesaw
mechanism becomes available. If lepton-number-violating mass terms are absent, the neutrino can instead
remain a Dirac fermion. What is an
optional distinction in the Standard Model therefore becomes a fork in
the realization of left--right symmetry.

The two routes lead to sharply different experimental questions. The
Majorana theory offers lepton-number violation, neutrinoless double beta
decay, and the production of heavy right-handed neutrinos. The Dirac route
removes these characteristic signals while retaining the parity relation
between the left- and right-handed flavor sectors. Parity
constrains the physical domain of the theory and connects the reconstruction
of right-handed mixing to spontaneous parity breaking and strong CP.

We begin by distinguishing these two realizations. The Majorana theory provides the setting for the
seesaw reconstruction program, while the Dirac theory will lead us to a
relation among leptonic CP violation, the absolute neutrino mass and
strong CP.
We also derive an invariant reduction of the Dirac compatibility
condition, together with new relations for the modulus mismatch
and relative CP orientation.

\section{The Majorana--Dirac fork}
\label{sec:fork}
Both realizations use the same fermion content and a scalar bidoublet
\(\Phi\sim(2,2,0)\), where the three entries refer to
\(SU(2)_L\times SU(2)_R\times U(1)_{B-L}\). The bidoublet couples
left- and right-handed fermions and generates their Dirac mass terms.
Generalized parity interchanges the two chiral sectors and sends
\(\Phi\) to \(\Phi^\dagger\), requiring the two bidoublet Yukawa
matrices in each fermion sector to be Hermitian~\cite{Senjanovic:1975rk}. This common flavor
structure survives the choice of fields responsible for breaking
\(SU(2)_R\).

In the conventional Majorana realization, this breaking is driven by
scalar triplets~\cite{Mohapatra:1980yp},
\begin{equation}
\Delta_L\sim(3,1,2),
\qquad
\Delta_R\sim(1,3,2).
\label{eq:triplets_review}
\end{equation}
Their gauge quantum numbers allow couplings to pairs of lepton doublets.
A neutral expectation value for \(\Delta_R\) therefore gives a Majorana
mass to the right-handed neutrinos at the same stage at which the new
right-handed gauge bosons become massive. An induced expectation value
for \(\Delta_L\) can contribute directly to the light-neutrino mass.
The two contributions are combined by the seesaw relation discussed
below.

The minimal Dirac realization instead uses scalar doublets~\cite{Senjanovic:1978ev},
\begin{equation}
\phi_L\sim(2,1,1),
\qquad
\phi_R\sim(1,2,1).
\label{eq:doublets_review}
\end{equation}
These fields cannot replace the triplets in a renormalizable coupling to
a pair of lepton doublets. With the minimal fermion content and
lepton-number conservation, the bidoublet then supplies the complete
neutrino mass. In referring to a Dirac theory, we assume that
lepton-number-violating higher-dimensional operators are also excluded.
The doublet choice by itself does not forbid such operators in a more
general effective theory.

These choices affect both the particles to be sought and the information
needed to test neutrino mass. The triplet theory relates light neutrinos
to a heavy Majorana sector and to scalars with lepton-number-violating
couplings. The minimal Dirac theory has three light neutrinos with
right-handed charged-current interactions, but no heavy Majorana mass
matrix to determine. Parity still constrains their flavor couplings.
The reconstruction problem consequently changes from recovering the
Dirac mass term inside the seesaw to determining the right-handed mixing
of the light Dirac neutrinos~\cite{Tello:2026ine}.

\section{The Majorana route}
\label{sec:majorana}
When the right-handed Majorana masses are large compared with the Dirac
entries, integrating out the heavy neutrinos gives
\begin{equation}
M_{\rm light}
=
M_L-M_D^T M_R^{-1}M_D .
\label{eq:seesaw_review}
\end{equation}
Here \(M_L\) and \(M_R\) are the symmetric Majorana mass matrices
generated by the triplets, and the placement of the transposes fixes our
convention for \(M_D\). The first term is the direct triplet, or type-II,
contribution~\cite{Mohapatra:1980yp}, while the second is the type-I
seesaw contribution~\cite{Minkowski:1977sc,Mohapatra:1979ia,Yanagida:1979as,Gell-Mann:1979vob,Glashow:1979nm}. Parity
relates the underlying left- and right-handed triplet Yukawa couplings.
It also constrains \(M_D\) through the Hermitian bidoublet Yukawas, so
these matrices belong to a common theory of symmetry breaking.

The experimental interest of this realization lies in the presence of heavy Majorana neutrinos that couple to \(W_R\) and can therefore be produced at colliders. The Keung--Senjanovi\'c process~\cite{Keung:1983uu} provides a particularly direct probe,
\begin{equation}
pp\longrightarrow W_R^\pm
\longrightarrow \ell_\alpha^\pm N_i
\longrightarrow \ell_\alpha^\pm\ell_\beta^\pm jj .
\label{eq:KS_review}
\end{equation}
The same-sign channel tests lepton-number violation, while the
reconstructed masses and the charged-lepton flavors probe the heavy
spectrum and its right-handed mixing~\cite{Das:2012ii,Vasquez:2014mxa,Cai:2026les}.
Its observability depends on the masses, branching fractions and heavy
neutrino lifetime. Prompt dilepton searches, displaced decays and missing
energy therefore address different parts of the same parameter space~\cite{Nemevsek:2018bbt}.
LHC searches in the KS channel constrain \(W_R\) masses in the
several-TeV range~\cite{ATLAS:2023cjo,CMS:2021dzb}.
Kaon mixing and CP violation provide complementary indirect
constraints on the right-handed scale~\cite{Bertolini:2019out,Dekens:2021bro}.

Neutrinoless double beta decay probes complementary combinations of
these couplings. Besides light-neutrino exchange, the LR theory admits
contributions involving right-handed currents, heavy neutrinos and
triplet scalars. The relation between a decay rate and a neutrino mass
therefore depends on which contributions dominate, their phases and the
nuclear matrix elements. The connection to collider measurements is
especially transparent in limits such as type-II dominance~\cite{Tello:2010am,Li:2020flq}.
Searches for charged-lepton flavor violation, such as
\(\mu\to e\gamma\) and \(\mu\to3e\), provide complementary
constraints on the heavy-neutrino and triplet-scalar flavor structure~\cite{Cirigliano:2004mv,Tello:2010am,Bonilla:2016fqd}.
Scalar studies provide further information about the origin of the mass
terms~\cite{Fuks:2025jrn}. Doubly charged scalars couple directly to lepton pairs, and
mixing of a neutral triplet scalar with the observed Higgs can allow
Higgs decays into heavy neutrinos when kinematically accessible~\cite{Senjanovic:2016vxw,Maiezza:2015lza}.

The reconstruction program uses this information to ask whether the
measured masses and couplings satisfy Eq.~\eqref{eq:seesaw_review}.
The light-neutrino mass matrix alone does not determine a general
\(M_D\)~\cite{Casas:2001sr}. Generalized parity constrains this freedom
through the Hermitian bidoublet Yukawas.\footnote{The alternative LR
symmetry, generalized charge conjugation, exchanges left-handed
multiplets with the charge conjugates of their right-handed partners
and requires a symmetric \(M_D\)~\cite{Nemevsek:2012iq}. Explicit
solutions have also been implemented in collider
models~\cite{Kriewald:2024cgr}.}
In the Hermitian limit, \(M_D\) can be
reconstructed from the light and heavy mass matrices and the triplet
contribution, subject to consistency conditions and discrete choices~\cite{Senjanovic:2019moe}.
Away from this limit, parity constrains its anti-Hermitian part.
For conventional seesaw Dirac masses, the very small deformation
allowed by strong CP gives only a tiny relative departure from
Hermiticity.
Heavy-neutrino decays into left-handed charged leptons~\cite{Nemevsek:2012iq,Helo:2018rll} and
scalar decays can then test the remaining Dirac Yukawa information~\cite{Senjanovic:2016vxw,Senjanovic:2018xtu}.
Numerical methods extend the reconstruction beyond the Hermitian
limit~\cite{Kiers:2022cyc}. The Hermitian solution has been used to
study radiative strong CP~\cite{Li:2024sln}. Further studies consider
quasi-degenerate heavy-neutrino spectra~\cite{Mikulenko:2024hex}, and
thermal leptogenesis, where relating the baryon asymmetry to
low-energy CP violation requires additional flavor
assumptions~\cite{Chen:2026jsc}.

Parity thus turns the seesaw into a constrained inverse problem
connecting light-neutrino data with the heavy and scalar sectors.
We now turn to the Dirac realization, where the heavy Majorana
sector is absent and parity relates the left- and right-handed
mixing matrices of the light neutrinos.

\section{The Dirac route}
\label{sec:dirac}
In the minimal doublet theory, the right-handed partners of the observed
neutrinos remain light. The characteristic Majorana signals discussed
above are absent under the assumption of lepton-number conservation.
The new gauge bosons and the extended scalar sector remain, however,
and searches for right-handed currents continue to test the theory.
For example, a \(W_R\) decay into a charged lepton and an unobserved
light neutrino gives missing energy rather than the heavy-neutrino decay
chain of Eq.~\eqref{eq:KS_review}. Inclusive rates sum over the
unobserved neutrino states and do not by themselves determine the
entries of \(V_R^\ell\).
LHC searches for dijet and top--bottom resonances also constrain
\(W_R\) in both realizations~\cite{Solera:2023kwt}.
The doublet realization is also constrained by electroweak
precision measurements~\cite{Bernard:2020cyi} and Higgs
data~\cite{Karmakar:2022iip}.

In a thermal cosmological history, right-handed interactions can also
populate the light \(\nu_R\) states, giving complementary constraints
through the effective number of relativistic species~\cite{Abazajian:2019oqj,Borah:2020boy}.

Parity nevertheless relates \(V_R^\ell\) to quantities in the
left-handed sector. This reconstruction follows the strategy developed
for right-handed quark mixing, from early analytic results to the more
general parity reconstruction~\cite{Zhang:2007fn,Senjanovic:2014pva,Senjanovic:2015yea}.
The much stronger neutrino--charged-lepton hierarchy leads to a
different physical domain and response~\cite{Tello:2026ine}.
To specify the conventions, write the bidoublet expectation value as
\(\langle\Phi\rangle=v\,\diag(\cos\beta,\sin\beta e^{i\alpha})\).
The Dirac mass matrices are
\begin{equation}
\begin{split}
M_\nu&=v\left(\cos\beta\,Y_1^\ell
             +\sin\beta\,e^{-i\alpha}Y_2^\ell\right),\\
M_\ell&=v\left(\sin\beta\,e^{i\alpha}Y_1^\ell
             +\cos\beta\,Y_2^\ell\right),
\end{split}
\label{eq:Dirac_masses_review}
\end{equation}
with \(Y_{1,2}^\ell=Y_{1,2}^{\ell\dagger}\). Arranging the observed
neutrino--charged-lepton hierarchy requires a suitable Yukawa and vacuum
structure. The reconstruction determines its flavor consequences without
by itself explaining the smallness of the Dirac masses.\footnote{Early work
examined the naturalness of Dirac neutrino masses in LR
theories~\cite{Branco:1978bz}. Small, calculable masses can arise in extended
LR realizations~\cite{Chang:1986bp}, including through two-loop
corrections~\cite{Babu:2022ikf}.} Their departure from Hermiticity is controlled by
\begin{equation}
\eps\equiv\sin\alpha\tan2\beta .
\label{eq:epsilon_review}
\end{equation}
This parameter measures the CP-odd bidoublet deformation relevant to
flavor. It can vanish while \(SU(2)_R\) remains broken.

Let \(M_f=U_L^{(f)}m_fU_R^{(f)\dagger}\), with \(m_f\) diagonal
and positive. The mismatch between its two diagonalizers is
\(U_f\equiv U_L^{(f)\dagger}U_R^{(f)}\). The mixing matrices obey
\begin{equation}
\begin{gathered}
V_L=U_L^{(\ell)\dagger}U_L^{(\nu)},
\qquad
V_R^\ell=U_R^{(\ell)\dagger}U_R^{(\nu)},\\
V_LU_\nu=U_\ell V_R^\ell .
\end{gathered}
\label{eq:mixing_mismatch_review}
\end{equation}
At \(\eps=0\), the mismatch matrices reduce to diagonal sign
matrices \(S_e\) and \(S_\nu\), giving
\(V_R^\ell=S_eV_LS_\nu\). These signs record the possible eigenvalue
signs of the Hermitian mass matrices before their physical masses are
made positive. They label reconstruction branches and are not negative
physical masses or Majorana phases. The corresponding signed masses are
\(\hat m_\alpha=(S_e)_{\alpha\alpha}m_\alpha\), for
\(\alpha=e,\mu,\tau\), and
\(\hat m_{\nu_i}=(S_\nu)_{ii}m_{\nu_i}\).

In the physical hierarchy \(m_\nu\ll m_\ell\), the charged-lepton
mismatch is \(U_\ell\simeq S_e\), up to negligible corrections and
diagonal phases that can be absorbed into the charged-lepton convention.
The coupled parity constraints then reduce to
\begin{equation}
X^2=m_\nu^2+i\eps HX,
\label{eq:master_review}
\end{equation}
where \(X\equiv U_\nu m_\nu\),
\(H\equiv V_L^\dagger\hat m_\ell V_L\), and
\(\hat m_\ell\equiv S_em_\ell\).
The derivation and the neglected hierarchy corrections are given in
Ref.~\refcite{Tello:2026ine}. We use this reduced system throughout the
following analysis. Exact identities on its physical solutions retain
the full nonlinear dependence on \(\eps\), while statements about the
underlying Yukawa matrices can be formulated before taking the
hierarchy limit.

An algebraic root of Eq.~\eqref{eq:master_review} must also represent a
unitary mismatch matrix. For nonzero neutrino masses the physical
conditions can be written as
\begin{equation}
X^\dagger X=m_\nu^2,
\qquad
X-X^\dagger=i\eps H .
\label{eq:U_master}
\end{equation}
The reconstructed mixing is then
\begin{equation}
V_R^P=S_eV_LXm_\nu^{-1},
\label{eq:VR_parity_review}
\end{equation}
where the superscript \(P\) indicates that only parity has been imposed.
For fixed neutrino masses, \(V_L\) and \(\eps\), one solves
Eq.~\eqref{eq:master_review} on each sign branch, retains the roots
satisfying Eq.~\eqref{eq:U_master}, and obtains \(V_R^P\) through
Eq.~\eqref{eq:VR_parity_review}.

Unitarity also limits the size of the deformation. In particular,
\begin{equation}
|\eps|\leq
\min_{i,j  }
\frac{m_{\nu_i}+m_{\nu_j}}{|H_{ij}|} .
\label{eq:unitarity_bound_review}
\end{equation}
This necessary bound is useful analytically, although finding the full
physical boundary requires solving the matrix problem. For generic
mixing, its scale is set by the lightest neutrino mass relative to the
charged-lepton masses. Opposite sign assignments bring neutrino mass
differences into the response denominators, allowing sizable mixing
corrections even for small \(\eps\).

At this stage the PMNS phase is still an input. Predicting it requires
an additional condition on the Yukawa sector. We shall impose
\(\CPell\), meaning that the two Dirac lepton Yukawa matrices admit a
common real parity basis. This restriction is imposed only in the lepton
sector, while the quark Yukawa matrices remain general Hermitian
matrices.
Compatibility with the reconstructed \(V_R^P\) will then
select the allowed leptonic phases~\cite{Tello:2026tun}.

\section{Leptonic reality as a compatibility condition}
\label{sec:compatibility}

Parity and leptonic reality constrain the theory in complementary ways.
Generalized parity determines the right-handed leptonic mixing matrix,
whereas \(\CPell\) specifies how this reconstructed matrix must be related
to the observed left-handed one. Imposing both requirements gives a
compatibility condition on the PMNS matrix.

\subsection{Reality, phases, and rephasing invariance}

The statement that a matrix is real is basis dependent. The physical
content of \(\CPell\) is that there exists a parity basis in which
the two Dirac lepton Yukawa matrices are simultaneously real. Since parity
requires
\begin{equation}
Y_{1,2}^{\ell}=Y_{1,2}^{\ell\dagger},
\end{equation}
such a basis makes them real symmetric.

After the bidoublet acquires its complex vacuum expectation value, the
charged-lepton and neutrino mass matrices are then complex symmetric. Their
left- and right-handed diagonalizers may consequently be chosen as
\begin{equation}
U_R^{(f)}=U_L^{(f)*}K_f,
\end{equation}
where \(K_f\) is diagonal and unitary. The phases in \(K_f\) reflect the
freedom to rephase Dirac mass eigenstates and contain no independent mixing
information. It follows that leptonic reality requires
\begin{equation}
V_R^\ell=K_e^\dagger V_L^*K_\nu .
\label{eq:CP_relation}
\end{equation}

Equation~\eqref{eq:CP_relation} expresses \(\CPell\) in mixing space,
with the matrices \(K_e\) and \(K_\nu\) accounting for
independent row and column rephasings. It immediately follows that
\begin{equation}
\left|V_R^\ell\right|^2=\left|V_L\right|^2,
\qquad
J_R=-J_L ,
\label{eq:moduli_J_requirements_review}
\end{equation}
where \(J_L\) and \(J_R\) denote the Jarlskog invariants of the two
mixing matrices. Complex conjugation reverses the orientation of CP
violation and hence the sign of the invariant.

A simple parameter count already indicates that these apparently numerous
conditions can encode only one physical restriction. A generic pair of
Hermitian \(3\times3\) Yukawa matrices contains \(18\) real parameters.
After removing the eight redundant parameters associated with a common
\(U(3)\) flavor transformation (the overall phase acts trivially), one
is left with ten physical parameters, comprising six masses, three mixing
angles, and one Dirac phase. A pair of real symmetric matrices contains
\(12\) parameters, and removing the freedom of a common \(O(3)\)
rotation leaves nine.
Thus \(\CPell\) removes precisely one physical parameter from the
parity-invariant lepton sector.

\subsection{One invariant obstruction}

The single restriction can be expressed at the Yukawa level through a
Jarlskog-type CP-odd invariant~\cite{Jarlskog:1985ht,Jarlskog:1985cw}.
For generic, nondegenerate three-generation Yukawas, define
\begin{equation}
\mathcal I_Y
\equiv
\frac{1}{6i}
\Tr\!\left([Y_1^\ell,Y_2^\ell]^3\right).
\label{eq:IY_review}
\end{equation}
In a basis in which
\(Y_1^\ell=\diag(y_1,y_2,y_3)\), this becomes
\begin{equation}
\mathcal I_Y
=
(y_1-y_2)(y_2-y_3)(y_3-y_1)
\Im\!\left[
(Y_2^\ell)_{12}(Y_2^\ell)_{23}(Y_2^\ell)_{31}
\right].
\end{equation}
Two diagonal rephasings make two off-diagonal entries of \(Y_2^\ell\)
real. The remaining entry can be made real if and only if their closed
product is real. Consequently,
\begin{equation}
\Pcal+\CPell\ \text{compatibility}
\quad\Longleftrightarrow\quad
\mathcal I_Y=0
\label{eq:IY_equivalence_review}
\end{equation}
for generic three-generation Yukawas. Degenerate spectra require the
usual separate treatment because of their enhanced flavor freedom.

More general weak-basis-invariant CP conditions for the LR fermion
mass sector were derived by Branco and Rebelo~\cite{Branco:1986pr}.

The same obstruction admits a particularly transparent form on the
physical parity solutions. Recall the mass-weighted neutrino mismatch
\(X=U_\nu m_\nu\), and define
\begin{equation}
G\equiv V_LXV_L^\dagger
=
\mathcal H+\frac{i\eps}{2}\hat m_\ell,
\qquad
\mathcal H\equiv
\frac12 V_L(X+X^\dagger)V_L^\dagger .
\label{eq:Hcore_review}
\end{equation}
The matrix \(\mathcal H\) is Hermitian, while the anti-Hermitian part of
\(G\) is diagonal. Hence \(G\) and \(\mathcal H\) have identical
off-diagonal entries.

Under a diagonal flavor rephasing,
\(\mathcal H_{\alpha\beta}\) acquires the phase difference between
flavors \(\alpha\) and \(\beta\). For three generations the phases cancel
around the closed cycle
\begin{equation}
\mathcal H_{e\mu}\mathcal H_{\mu\tau}\mathcal H_{\tau e}.
\end{equation}
Its imaginary part,
\begin{equation}
\mathcal I_{\mathcal H}
\equiv
\Im\!\left(
\mathcal H_{e\mu}
\mathcal H_{\mu\tau}
\mathcal H_{\tau e}
\right),
\label{eq:IH_review}
\end{equation}
is therefore rephasing invariant. Two off-diagonal phases of
\(\mathcal H\) can always be removed. The third can be removed precisely
when \(\mathcal I_{\mathcal H}=0\). In that basis \(\mathcal H\) is real
symmetric and \(G\) is complex symmetric. Its left and right singular
vectors are then related by complex conjugation, yielding
Eq.~\eqref{eq:CP_relation}. Thus, on the physical parity
image,
\begin{equation}
\Pcal+\CPell\ \text{compatibility}
\quad\Longleftrightarrow\quad
\mathcal I_{\mathcal H}=0 .
\label{eq:IH_equivalence_review}
\end{equation}
The proof and the qualifications at exceptional points are discussed in
Appendix~\ref{app:exact_scalar_compatibility}.

\subsection{A single amplitude for the modulus mismatch}

The same condition constrains the left--right mixing differences. Introduce
\begin{equation}
L\equiv G^\dagger G
=
V_Lm_\nu^2V_L^\dagger,
\qquad
R\equiv GG^\dagger
=
S_eV_R^P m_\nu^2V_R^{P\dagger}S_e ,
\label{eq:LR_review}
\end{equation}
and define the right--left modulus mismatch
\begin{equation}
\mathcal M_{\alpha i}
\equiv
\left|(V_R^P)_{\alpha i}\right|^2
-
\left|(V_L)_{\alpha i}\right|^2 .
\label{eq:Mmod_review}
\end{equation}

Unitarity and equality of the diagonal elements of \(L\) and \(R\) give,
for every charged-lepton flavor,
\begin{equation}
\sum_i\mathcal M_{\alpha i}=0,
\qquad
\sum_i m_{\nu_i}^2\mathcal M_{\alpha i}=0.
\label{eq:row_sumrules_review}
\end{equation}
For nondegenerate neutrino masses, each row must therefore be proportional
to
\begin{equation}
\boldsymbol{\mathcal N}
=
\begin{pmatrix}
m_{\nu_3}^2-m_{\nu_2}^2\\
m_{\nu_1}^2-m_{\nu_3}^2\\
m_{\nu_2}^2-m_{\nu_1}^2
\end{pmatrix}.
\label{eq:N_review}
\end{equation}

The master equation, Eq.~\eqref{eq:master_review}, supplies the
corresponding column constraints,
\begin{equation}
\sum_\alpha\mathcal M_{\alpha i}=0,
\qquad
\sum_\alpha\hat m_\alpha\mathcal M_{\alpha i}=0.
\label{eq:column_sumrules_review}
\end{equation}
Every column is consequently proportional to
\begin{equation}
\boldsymbol{\mathcal E}
=
\begin{pmatrix}
\hat m_\mu-\hat m_\tau\\
\hat m_\tau-\hat m_e\\
\hat m_e-\hat m_\mu
\end{pmatrix}.
\label{eq:E_review}
\end{equation}
The complete \(3\times3\) mismatch must therefore take the outer-product
form
\begin{equation}
\mathcal M
=
\Lambda\,
\boldsymbol{\mathcal E}
\boldsymbol{\mathcal N}^{T}
\label{eq:M_outer_product_review}
\end{equation}
for a single real amplitude \(\Lambda\), with
\(\operatorname{rank}\mathcal M\leq1\).
Thus all nine modulus differences are governed by a single amplitude.
For each physical root, \(\Lambda\) is determined by the masses, the
left-handed mixing angles and phase, \(\eps\), and the sign branch.
This is a special feature of the parity reconstruction. On compatible
solutions \(\mathcal M=0\).

Defining
\begin{equation}
\mathcal D_\nu
\equiv
(m_{\nu_1}^2-m_{\nu_2}^2)
(m_{\nu_2}^2-m_{\nu_3}^2)
(m_{\nu_3}^2-m_{\nu_1}^2),
\label{eq:Dnu_review}
\end{equation}
the next spectral moment
\(\bigl(\sum_i m_{\nu_i}^4\mathcal M_{\alpha i}\bigr)\)
gives the exact relation
\begin{equation}
\Lambda\mathcal D_\nu
=
2\eps\,\mathcal I_{\mathcal H}.
\label{eq:Lambda_IH_review}
\end{equation}
For \(\eps\neq0\) and nondegenerate neutrino masses, modulus alignment,
\(\Lambda=0\), is therefore already equivalent to full compatibility. At
\(\eps=0\), parity makes the moduli aligned identically, and
the invariant \(\mathcal I_{\mathcal H}\) remains the appropriate test.

\subsection{The Jarlskog projection}

The CP orientation provides a second projection of the same obstruction.
The exact three-generation algebra gives
\begin{equation}
(J_R+J_L)\mathcal D_\nu
=
2\mathcal I_{\mathcal H}\,\Xi ,
\label{eq:Jsum_IH_review}
\end{equation}
where
\begin{equation}
\Xi
\equiv
\det\!\left[
(\Tr\mathcal H)\mathbf1-\mathcal H
\right]
+
\frac{\eps^2}{4}
\sum_{\alpha=e,\mu,\tau}
\mathcal E_\alpha^2\mathcal H_{\alpha\alpha}.
\label{eq:Xi_review}
\end{equation}
The modulus mismatch and the relative CP orientation are thus two
observable projections of \(\mathcal I_{\mathcal H}\), as expressed by
Eqs.~\eqref{eq:Lambda_IH_review} and \eqref{eq:Jsum_IH_review}.

In the Hermitian limit, \(\eps=0\),
\begin{equation}
\Xi(0)
=
\prod_{i<j}
(\hat m_{\nu_i}+\hat m_{\nu_j}).
\label{eq:Xi_seed_review}
\end{equation}
Away from an exact signed degeneracy, \(\Xi\neq0\) in a neighborhood of
either CP-conserving solution. In this region,
\begin{equation}
J_R=-J_L
\quad\Longleftrightarrow\quad
\mathcal I_{\mathcal H}=0.
\label{eq:J_equivalence_review}
\end{equation}
The Jarlskog condition is therefore locally equivalent to the full
compatibility condition.
Near a zero of \(\Xi\), the full invariant condition must be used.

Finally, write the Jarlskog invariant reconstructed from parity alone as
\begin{equation}
J_R=J_L+\Delta J_P.
\end{equation}
Compatibility then gives the exact observable relation
\begin{equation}
J_L=-\frac12\,\Delta J_P .
\label{eq:J_projection_review}
\end{equation}
Parity computes the right-handed displacement \(\Delta J_P\), while
leptonic reality requires the observable left-handed invariant to supply
exactly half of it with the opposite sign. We use this relation to derive
the leading CP response.

\section{Compressed spectra and the strong-CP map}
\label{sec:response_triangle}

Equation~\eqref{eq:J_projection_review} turns the parity reconstruction
into a prediction for observable leptonic CP violation. The leading
response shows how a small parity deformation can generate large
leptonic CP violation.

At \(\eps=0\), compatibility requires
\(V_L\sim V_L^*\), and the branches originate from the two
CP-conserving points
\begin{equation}
\dCP=\delta_0,
\qquad
\delta_0=0,\pi .
\end{equation}
For the expansion near either origin, take
\(O\equiv V_L(\delta_0)\in SO(3)\). Expanding the parity reconstruction
to first order gives
\begin{equation}
J_{\rm CP}^{\ell}
=
-\frac{\eps}{2}
\sum_{i<j}
\frac{
(O^T\hat m_\ell O)_{ij}
}{
\hat m_{\nu_i}+\hat m_{\nu_j}
}
\,O_{ek}O_{\mu k}O_{\tau k}
+\mathcal O(\eps^3),
\qquad
\{i,j,k\}=\{1,2,3\}.
\label{eq:J_response_review}
\end{equation}
The absence of even powers follows from conjugation symmetry. The
CP-conjugate branch maps
\(\eps\to-\eps\) and
\(J_{\rm CP}^{\ell}\to-J_{\rm CP}^{\ell}\).

Equation~\eqref{eq:J_response_review} separates the three neutrino-pair
response channels. For equal neutrino parity signs, the denominators are
ordinary mass sums, without enhancement from small mass differences. For opposite
signs, a sum becomes a mass difference. The small oscillation splittings
can then amplify the effect of a tiny parity deformation.

For example, keeping the enhanced \(12\) contribution on a branch with
\(S_\nu=(+,-,\pm)\) gives
\begin{equation}
\left.\dCP\right|_{12}\simeq\delta_0
+
\frac{\eps\hat m_\tau}
{2(m_{\nu_1}-m_{\nu_2})}
C_{12}^{\tau}
\label{eq:eta12_review}
\end{equation}
where the subscript \(12\) denotes the approximation retaining that
response channel. The angular coefficient \(C_{12}^{\tau}\) depends on
the measured mixing angles and on the choice of \(\delta_0\). Explicit
expressions for the angular coefficients are given in
Ref.~\refcite{Tello:2026tun}. Here \(J_{\max}=c_{12}s_{12}c_{23}s_{23}c_{13}^2s_{13}\),
\(s_{ij}=\sin\theta_{ij}\) and \(c_{ij}=\cos\theta_{ij}\), so that
\(J_{\rm CP}^{\ell}=J_{\max}\sin\dCP\).
The charged-lepton hierarchy makes the tau contribution dominant,
while the small solar splitting controls the enhancement. The response
therefore grows as the neutrino spectrum becomes compressed.

Once the ordinary expansion ceases to be controlled, the response must
be obtained from the nonlinear parity equation or from a different
controlled expansion.

In the quasi-degenerate regime one writes
\begin{equation}
m_{\nu_i}^2=m_0^2(1+\Delta_i),
\qquad
|\Delta_i|\ll1 ,
\end{equation}
and resums the dependence on \(\eps m_\alpha/m_0\) into charged-lepton
phase factors. For the QD continuation of the representative atmospheric
branch analyzed in Appendix~A of Ref.~\refcite{Tello:2026tun}, the leading
response near
\(\delta_0\), further expanded for \(|\eps|m_\tau/(2m_0)\ll1\), is
\begin{equation}
\left.\dCP\right|_{\rm QD}\simeq \delta_0
+
\frac{\Delta m_{31}^2}
{\eps\hat m_\tau m_0}
C_{12}^{\tau}.
\label{eq:etaQD_review}
\end{equation}
The omitted term proportional to
\(\Delta m_{21}^2C_{13}^{\tau}\) changes the leading atmospheric result
by about ten percent for the benchmark mixing angles.

Both expressions describe a small displacement from \(\delta_0\) in
their respective domains of validity. The inverse dependence on
\(\eps\) in Eq.~\eqref{eq:etaQD_review} cannot be extrapolated to
\(\eps=0\). Between the controlled limits, the response can become
large and must be obtained from the nonlinear compatibility condition.
This is how a deformation small enough to satisfy the strong-CP bound
can still generate sizable, and on some branches maximal, leptonic CP
violation. The full numerical solutions are given in
Ref.~\refcite{Tello:2026tun}.

\subsection{From parity breaking to strong CP}

The physical strong-CP angle is
\(\bar\theta=\theta_{\rm QCD}+\arg\det(M_uM_d)\).
Parity restricts the bare QCD angle to \(0\) or \(\pi\)~\cite{Beg:1978mt,Mohapatra:1978fy,Barr:1991qx}.
We choose the parity branch with \(\theta_{\rm QCD}=0\) and quark sign
assignments for which \(\bar\theta(0)=0\). The complex bidoublet
expectation value then induces the determinant phase through the
departure of the quark mass matrices from Hermiticity, controlled by
the same parameter \(\eps\). On a matched quark parity branch
\(S=(S_u,S_d)\), the induced strong-CP phase is
\begin{equation}
\bar\theta_S
=
C_q^{(S)}\eps+\mathcal O(\eps^2),
\label{eq:bartheta_review}
\end{equation}
with
\begin{equation}
C_q^{(S)}
=
\frac12
\sum_{i,j}
s_{u_i}s_{d_j}
\left|(V_{\rm CKM})_{ij}\right|^2
\left(
\frac{m_{u_i}}{m_{d_j}}
-
\frac{m_{d_j}}{m_{u_i}}
\right).
\label{eq:Cq_review}
\end{equation}
The quark masses and CKM matrix entering \(C_q^{(S)}\) are evaluated at
a common parity matching scale.
The familiar scaling
\(\bar\theta\sim(m_t/2m_b)\eps\) corresponds to the dominant term
on the conventional branch~\cite{Maiezza:2014ala,Senjanovic:2015yea,Bertolini:2019out,Dekens:2021bro}.
Equation~\eqref{eq:Cq_review} keeps the full flavor and sign
dependence in the conventions of Ref.~\refcite{Tello:2026tun}.

We assume here that no Peccei--Quinn relaxation~\cite{Peccei:1977hh} or
accidental cancellation removes the parity-induced contribution. The
neutron-EDM constraint then sets a branch-dependent upper limit on
\(|\eps|\). Independently, the unitarity bound in
Eq.~\eqref{eq:unitarity_bound_review} gives the characteristic scale
of the physical leptonic parity domain,
\begin{equation}
|\eps|_{\max}
\sim
\frac{m_{\rm lightest}}{m_\tau},
\label{eq:eps_domain_review}
\end{equation}
up to mixing and branch factors. The absolute neutrino mass therefore
controls both the existence of the reconstruction and the amount of parity
breaking available to generate leptonic CP violation.

\subsection{The observable triangle}

For fixed oscillation inputs, mass ordering, and leptonic branch
\(\mathcal B_\ell\), the compatibility solutions define
\begin{equation}
\frac{J_{\rm CP}^{\ell}}{J_{\max}}
=
F_\ell\!\left(
m_{\rm lightest},|\eps|;
\mathcal B_\ell,\text{ordering}
\right).
\label{eq:Fell_review}
\end{equation}
The branch label retains the sign choices and the CP orientation when
\(|\eps|\) is used. If several roots occur, each is treated as a separate
local branch. Combining this relation with the quark reconstruction gives
\begin{equation}
\frac{J_{\rm CP}^{\ell}}{J_{\max}}
\simeq
F_\ell\!\left(
m_{\rm lightest},
\frac{|\bar\theta|}{|C_q^{(S)}|};
\mathcal B_\ell,\text{ordering}
\right).
\label{eq:triangle_review}
\end{equation}

Equation~\eqref{eq:triangle_review} defines a family of branch-labelled
surfaces in the observable space
\begin{equation}
\left(
J_{\rm CP}^{\ell},
m_{\rm lightest},
|\bar\theta|
\right).
\end{equation}
Oscillation experiments determine the mixing angles and constrain the
leptonic phase~\cite{Esteban:2024eli,NuFIT:2025v61}.
DUNE, Hyper-Kamiokande and the proposed ESS\(\nu\)SB experiment
aim to measure the leptonic phase more precisely~\cite{DUNE:2020jqi,Hyper-Kamiokande:2025fci,ESSnuSB:2023ogw}.
Direct beta-decay experiments such as KATRIN probe
\(m_\beta^2=\sum_i |V_{ei}|^2m_{\nu_i}^2\), which can be converted
into an absolute mass scale once the ordering and splittings are
given~\cite{KATRIN:2024cdt}.
  Project~8 pursues improved direct mass sensitivity~\cite{Project8:2022wqh}.
Cosmological information on \(\sum_i m_{\nu_i}\) provides an additional,
cosmology-dependent constraint~\cite{Elbers:2025vlz}. Neutron-EDM searches
constrain \(\bar\theta\) under the assumed dominance of its
contribution~\cite{Abel:2020pzs,Pospelov:2005pr,Engel:2013lsa,Ramsey-Musolf:2020ndm}.
  The n2EDM experiment aims to improve the experimental sensitivity~\cite{n2EDM:2021yah}.

A determination of \(J_{\rm CP}^{\ell}\) and \(m_{\rm lightest}\), for
example, selects compatible values of \(|\eps|\) and thus a
branch-dependent range of \(|\bar\theta|\). Conversely, a neutron-EDM
signal attributable predominantly to \(\bar\theta\), combined with the
oscillation measurement, would select intervals of the absolute neutrino
mass. Finite experimental uncertainties broaden the allowed regions, whereas
discrete branch choices and multiple nonlinear solutions can persist
even for exact inputs.

\begin{figure}[t]
\centering
\includegraphics[width=0.9\linewidth]{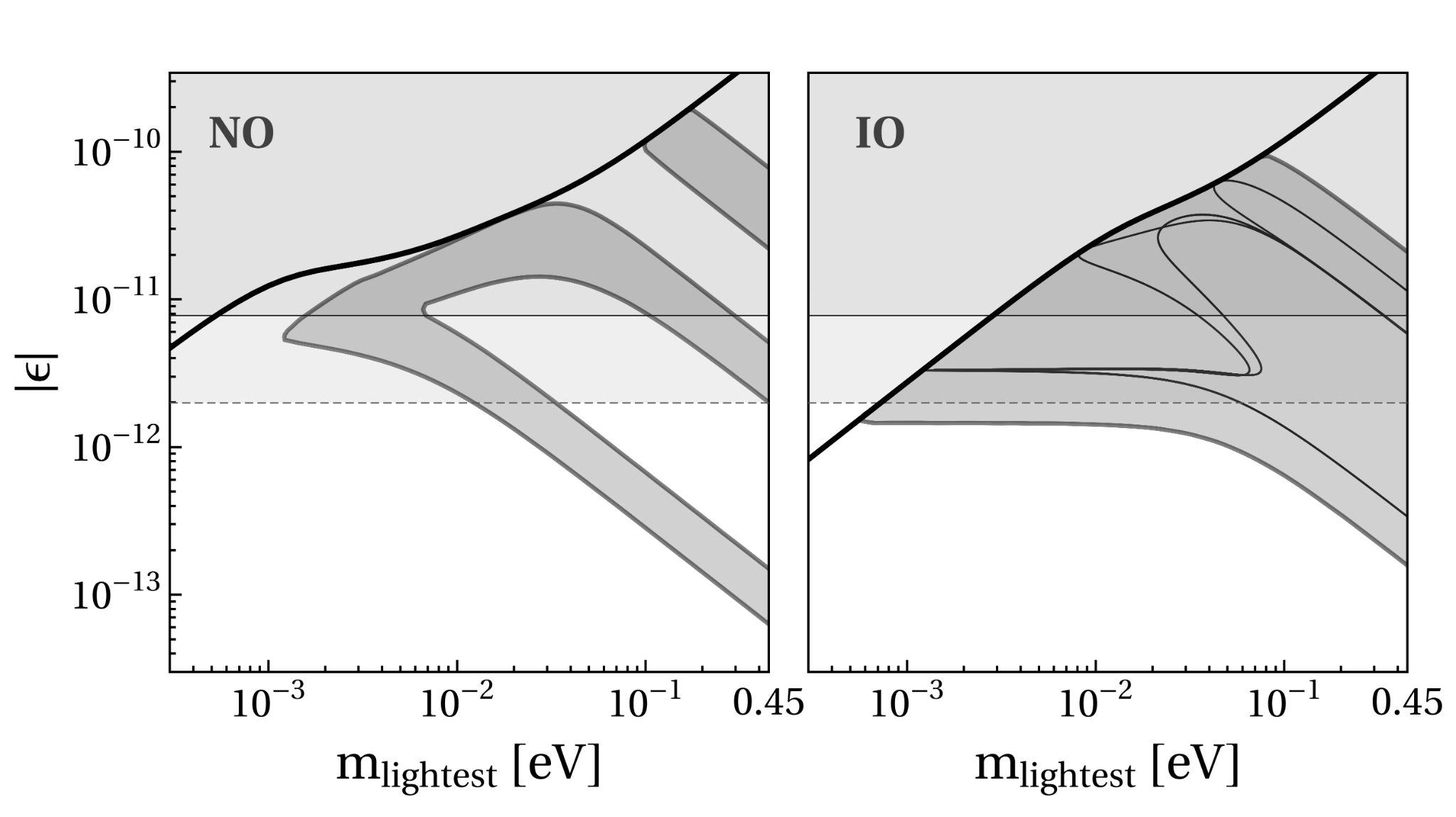}
\caption{Benchmark compatibility regions in the
\((m_{\rm lightest},|\eps|)\) plane for NO and IO. The shaded regions are
the union of all leptonic compatibility solutions satisfying
\(\dCP\in[200^\circ,225^\circ]\) in NO and
\(\dCP\in[250^\circ,290^\circ]\) in IO. In the IO panel, the thin black
curves show the exact maximal-CP contours
\(\dCP=90^\circ\) and \(270^\circ\), obtained without imposing the phase
windows. The thick black curve marks the upper envelope of the physical
compatibility domain with \(\dCP\) unrestricted. The horizontal dashed
and solid lines denote \(|\eps|=2\times10^{-12}\) and
\(7.8\times10^{-12}\). Below the dashed line
all matched quark parity
branches satisfy the nominal neutron-EDM constraint, between the two
lines only a subset does, and above the solid line none does.}
\label{fig:compatibility_bands_review}
\end{figure}

Figure~\ref{fig:compatibility_bands_review} illustrates this projection
using the solutions of Ref.~\refcite{Tello:2026tun} and the phase
windows specified in the caption. The shaded regions are illustrative
compatibility regions rather than statistical confidence regions.

The strong-CP boundaries use quark inputs evaluated at \(10~{\rm TeV}\)
and the nominal criterion \(|\bar\theta|<10^{-10}\).
The leptonic curves use central low-energy inputs without RG evolution~\cite{Lindner:2005as}.

\FloatBarrier

The maximal-CP contours show that several compatibility curves can
correspond to the same phase. A finite phase window broadens the
corresponding curves into regions. These regions are wider where the
predicted phase varies slowly with \(m_{\rm lightest}\) and \(|\eps|\),
and narrower where it varies rapidly.

The correlations must therefore be tested jointly for each allowed
pairing of leptonic and quark branches. Compatibility with the
measurements requires at least one common solution.

\section{Predicting leptonic CP: relation to other approaches}
\label{sec:literature_position}
The prospect of determining the leptonic Dirac phase from symmetry has
generated a large and varied model-building literature. Texture zeros,
mass sum rules, discrete flavor groups, residual generalized-CP
symmetries and modular symmetries reduce the number of independent
parameters in the lepton mass matrices and can give discrete values or
correlations for \(\dCP\)~\cite{King:2013eh,Feruglio:2019ybq,Chauhan:2023faf}.
Correlations with the absolute neutrino-mass scale also occur in
flavor-symmetric Dirac-neutrino models and in constructions imposing
quark--lepton relations on the Yukawa matrices~\cite{Bonilla:2017ekt,Haba:2018xke}.
These approaches provide useful examples of how information beyond the
measured masses and mixing angles can constrain leptonic CP violation.

A related program seeks a common origin for quark, leptonic and strong
CP. A spontaneously generated phase can enter both the CKM and PMNS
sectors while a Nelson--Barr~\cite{Nelson:1983zb,Barr:1984qx} structure
protects strong CP~\cite{Branco:2003rt}.
More recent axionless flavor constructions constrain the PMNS phase,
with narrow predictions in some realizations~\cite{Tanimoto:2024nwm,Tanimoto:2025fnj,Kobayashi:2025rpx}.
The resulting relations depend on how CP breaking is communicated to the
fermion sectors and on the flavor structure of the ultraviolet theory.

Within LR symmetry, the connection between parity and CP has a long
history. Early work analyzed the allowed CP-violating phases~\cite{Mohapatra:1978cr},
and subsequent studies developed spontaneous CP violation in minimal
\(\Pcal+\CP\) realizations~\cite{Branco:1982wp,Chang:1982dp,Ecker:1983hz}.
When the bidoublet Yukawa matrices are real in both the quark and lepton
sectors, the complex bidoublet expectation value supplies their CP
violation. In the usual minimal realization, the CKM phase is then tied
to the same vacuum deformation that induces strong CP. Reconciling the
observed quark phase with a very small \(\bar\theta\) is consequently
a central constraint on this construction.

Kuchimanchi's axionless \(\Pcal+\CP\) construction separates these
roles by placing CP breaking in a parity-conserving sector and
communicating it to a vectorlike quark family~\cite{Kuchimanchi:2010xs,Kuchimanchi:2012te}.
This generates a CKM phase without a tree-level strong-CP phase. In the
minimal construction there is no analogous vectorlike lepton family,
and the leptonic Dirac and Majorana phases vanish modulo \(\pi\) at
tree level. The radiatively induced leptonic phases were conservatively
estimated to remain very small, within approximately \(0.5^\circ\) of their
CP-conserving values. This estimate illustrates how separating the
communication of CP breaking to quarks and leptons can lead to a
testable expectation for the PMNS phase.

A further connection arises through radiative corrections in the
conventional triplet LR theory. Kuchimanchi showed that leptonic Yukawa
phases generate an imaginary part of the CP-sensitive scalar coupling
\(\alpha_2\)~\cite{Kuchimanchi:2014ota}.
Its one-loop flavor structure is schematically
\begin{equation}
\Im\alpha_2
\propto
\frac{i}{16\pi^2}
\Tr\!\left[f^\dagger f\,[h^\ell,\widetilde h^\ell]\right],
\label{eq:Kuchimanchi_invariant_review}
\end{equation}
where \(f\) is the Majorana triplet Yukawa coupling and
\(h^\ell,\widetilde h^\ell\) are the two Hermitian bidoublet lepton
Yukawa matrices. The proportionality sign leaves the running logarithm and
convention-dependent normalization implicit. The loop contribution
depends on the relative flavor structure and
can vanish for aligned configurations even when leptonic CP is
nonzero~\cite{deVries:2022nyh,Li:2024sln}. Bounds inferred from this
contribution are therefore flavor dependent. The induced scalar phase feeds
into \(\bar\theta\), so the neutron-EDM bound constrains leptonic CP
under assumptions about the Dirac Yukawas, the seesaw regime and the
absence of an axion or cancellations. In substantial regions of a recent
chiral realization, the leptonic phase is expected to remain within
approximately one degree of a CP-conserving value~\cite{Kuchimanchi:2024nkt}.

Babu, Mohapatra and Thapa obtain a different, particularly sharp
prediction in their \(SU(5)_L\times SU(5)_R\) construction~\cite{Babu:2023dzz}.
Parity provides a solution to strong CP, while a type-II Dirac-seesaw
mechanism and unified fermion-mass relations select normal ordering.
Their fermion-mass fit gives, at \(2\sigma\),
\begin{equation}
\begin{split}
\delta_{\rm CP}&=(130.4\pm1.2)^\circ
\quad\text{or}\quad(229.6\pm1.2)^\circ,\\
m_{\nu_1}&=4.8\text{--}8.4~{\rm meV}.
\end{split}
\label{eq:BMT_prediction_review}
\end{equation}

The Dirac compatibility approach developed above rests instead on
generalized parity, the minimal Dirac mass structure and a common real
basis for the two lepton Yukawa matrices. Parity first reconstructs
\(V_R^P\), and leptonic reality then requires it to be
rephasing-equivalent to \(V_L^*\). This gives a consistency condition
among observables without fixing a texture for the PMNS matrix.
The quark Yukawas remain general Hermitian matrices, so
the observed CKM phase is independent of \(\eps\). Also, the triplet
coupling \(f\) is absent, and the particular one-loop contribution in
Eq.~\eqref{eq:Kuchimanchi_invariant_review} does not arise. The connection
to strong CP used here comes from the quark-sector response to
the same bidoublet deformation.

Combining the two sectors gives a relation of the form
\begin{equation}
{\cal F}_{\mathcal B_\ell,S}
\!\left(J_{\rm CP}^{\ell},m_{\rm lightest},|\bar\theta|\right)=0,
\end{equation}
where \(\mathcal B_\ell\) labels the leptonic solution and \(S\) the
matched quark branch. The allowed phase is determined through this
relation once the remaining inputs and branches have been specified.
Predictions such as Eq.~\eqref{eq:BMT_prediction_review} provide useful
benchmarks for comparison with the allowed regions.

\section{Outlook}

\label{sec:outlook}
Left--right symmetry connects the existence of the right-handed neutrino
to the origin of parity violation, while allowing different realizations
of its mass. In the Majorana theory, the heavy neutrinos and triplet
scalars provide ways of testing lepton-number violation and reconstructing
the mass terms entering the seesaw. The Dirac realization changes the
experimental questions. Although the characteristic Majorana signals
are absent, parity still constrains right-handed flavor. The small Dirac
masses restrict the allowed parity deformation and hence its contribution
to strong CP.

The additional assumption of a common real parity basis for the lepton
Yukawa matrices makes this reconstruction predictive for leptonic CP.
Its compatibility condition has a simple physical content, despite the
several forms in which it can be expressed. For three generations, one
invariant controls both the modulus mismatch and the CP orientation.
Together with the quark-sector response to the same bidoublet
deformation, this gives correlations among the leptonic phase, the
absolute neutrino mass and strong CP. Compressed mixed-sign spectra
allow a small deformation to produce a large leptonic response, so a
stringent strong-CP constraint need not imply a nearly CP-conserving
PMNS matrix.

The phenomenological task is to confront these correlations with joint
information from oscillations, absolute-mass measurements and electric
dipole moments. Experimental uncertainties and the discrete branches
must both be retained in this comparison. The construction is excluded if no admissible pairing of leptonic and
quark branches accommodates the measured phase and absolute mass within
the strong-CP bound.

Several theoretical questions remain open. Running and threshold
matching are needed to connect a condition imposed at the parity scale
to precision low-energy data.
A completion must explain the lepton-sector reality condition and its
radiative stability in the presence of CP violation in the quark sector.
The scalar vacuum
and Yukawa structure must also account for the small Dirac masses and
the required bidoublet deformation. Such a completion could further
restrict the correlations and determine corrections to the tree-level
relation. The
reconstruction provides a way of testing parity through neutrino flavor
even when lepton number is conserved.

\begin{acknowledgments}
This review is based in part on a talk presented at LHC Days in Split 2026. I thank Goran Senjanovi\'c for discussions over the years on left-right symmetry, parity, and neutrino mass.
\end{acknowledgments}

\appendix
\numberwithin{equation}{section}

\section{Exact reduction of the compatibility condition}
\label{app:exact_scalar_compatibility}

In this appendix we show that, on the physical parity solutions, the full
\(\Pcal+\CPell\) compatibility requirement reduces to the vanishing of a
single scalar invariant of the Hermitian core \(\mathcal H\) introduced in
Eq.~\eqref{eq:Hcore_review}.

We first establish the compatibility condition directly. We then show how
the same scalar obstruction appears in the mixing moduli and in the
Jarlskog invariants.

For compactness, we introduce
\begin{equation}
G
\equiv
V_LXV_L^\dagger
=
\mathcal H+ \frac{i\eps}{2} \hat m_\ell
=
S_eV_R^Pm_\nu V_L^\dagger .
\label{eq:G_appendix}
\end{equation}
Since \(i \eps \hat m_\ell /2\) is diagonal, \(G\) and \(\mathcal H\) have
identical off-diagonal entries.

The corresponding mass-squared matrices \(L\) and \(R\), together with
the modulus mismatch \(\mathcal M\), are defined in
Eqs.~\eqref{eq:LR_review} and \eqref{eq:Mmod_review}.

\subsection{The scalar compatibility condition}

Under diagonal flavor rephasings, the closed product defining
\(\mathcal I_{\mathcal H}\) in Eq.~\eqref{eq:IH_review} is invariant. For
a generic \(3\times3\) Hermitian matrix it contains the only independent
CP-odd phase. We now prove the equivalence in
Eq.~\eqref{eq:IH_equivalence_review} for generic three-generation mixing
and a nondegenerate neutrino spectrum.

First assume compatibility in the form
\(V_R^P=K_e^\dagger V_L^*K_\nu\).
Equation~\eqref{eq:G_appendix} may then be written as
\begin{equation}
G
=
\Omega_e A,
\qquad
\Omega_e\equiv S_eK_e^\dagger ,
\qquad
A\equiv V_L^*K_\nu m_\nu V_L^\dagger ,
\label{eq:G_OmegaA}
\end{equation}
where
\begin{equation}
A^T=A .
\end{equation}
Writing
\(\Omega_e=\diag(\omega_e,\omega_\mu,\omega_\tau)\), one has
\begin{equation}
G_{\alpha\beta}
=
\omega_\alpha A_{\alpha\beta}.
\end{equation}
Parity, on the other hand, implies from Eq.~\eqref{eq:G_appendix} that the
off-diagonal part of \(G\) is Hermitian,
\begin{equation}
G_{\beta\alpha}=G_{\alpha\beta}^* .
\end{equation}
Therefore
\begin{equation}
G_{\alpha\beta}^*
=
\frac{\omega_\beta}{\omega_\alpha}
G_{\alpha\beta}.
\label{eq:G_phase_relation}
\end{equation}
Multiplying around the closed flavor cycle,
\begin{align}
\left(
G_{e\mu}G_{\mu\tau}G_{\tau e}
\right)^*
&=
\frac{\omega_\mu}{\omega_e}
\frac{\omega_\tau}{\omega_\mu}
\frac{\omega_e}{\omega_\tau}
G_{e\mu}G_{\mu\tau}G_{\tau e}
\nonumber\\
&=
G_{e\mu}G_{\mu\tau}G_{\tau e}.
\end{align}
The closed product is therefore real. Since \(G\) and \(\mathcal H\) have
the same off-diagonal entries,
\begin{equation}
\mathcal I_{\mathcal H}=0 .
\label{eq:IH_zero_appendix}
\end{equation}

Conversely, suppose that Eq.~\eqref{eq:IH_zero_appendix} holds. A diagonal
flavor rephasing can then make \(\mathcal H\) real. For generic
nonzero off-diagonal entries, two phases may be removed independently, and
reality of the closed product makes the third entry real as well. If one
off-diagonal entry vanishes, the conclusion requires fewer phase
conditions and follows directly.

To make the basis change explicit, let
\begin{equation}
P=\diag(e^{i\phi_e},e^{i\phi_\mu},e^{i\phi_\tau})
\end{equation}
and rephase
\begin{equation}
V_L'=PV_L,
\qquad
V_R'=PV_R^P .
\label{eq:flavor_rephasing}
\end{equation}
The relative rotation \(U_\nu=V_L^\dagger S_eV_R^P\), and therefore
\(X=U_\nu m_\nu\), are unchanged. Since \(P\) commutes with
\(S_e\) and \(\hat m_\ell\),
\begin{equation}
G'=PGP^\dagger,
\qquad
\mathcal H'=P\mathcal HP^\dagger .
\end{equation}
Choose \(P\) such that \(\mathcal H'\) is real. Since it remains
Hermitian, it is real symmetric, and
\begin{equation}
G'
=
\mathcal H'+\frac{i\eps}{2}\hat m_\ell
\end{equation}
is therefore complex symmetric,
\begin{equation}
G'^T=G'.
\label{eq:G_symmetric}
\end{equation}

At the same time,
\begin{equation}
G'
=
S_eV_R'm_\nu V_L'^\dagger .
\label{eq:Gprime_SVD}
\end{equation}
Since \(G'\) is complex symmetric, the same Takagi-type argument used in
deriving Eq.~\eqref{eq:CP_relation} relates its two unitary
diagonalizers by complex conjugation. For nondegenerate \(m_\nu\),
Eq.~\eqref{eq:Gprime_SVD} therefore implies
\begin{equation}
V_R'
=
S_eV_L'^*K_\nu'
\end{equation}
for a diagonal unitary \(K_\nu'\). Undoing the flavor rephasing gives
\begin{equation}
V_R^P
=
P^\dagger S_eP^*V_L^*K_\nu' .
\end{equation}
Since \(P^\dagger S_eP^*\) is diagonal and unitary, it may be identified
with \(K_e^\dagger\). Thus
\begin{equation}
V_R^P
=
K_e^\dagger V_L^*K_\nu ,
\end{equation}
which proves compatibility.

\subsection{Modulus projection and outer-product structure}

We next determine how the same scalar obstruction appears in the mixing
moduli. From Eq.~\eqref{eq:G_appendix},
\begin{align}
L
&=
\mathcal H^2+\frac{\eps^2}{4}\hat m_\ell^2
-\frac{ i\eps}{2}[\hat m_\ell,\mathcal H],
\label{eq:L_Hcore_expanded}
\\
R
&=
\mathcal H^2+\frac{\eps^2}{4}\hat m_\ell^2
+\frac{ i\eps}{2}[\hat m_\ell,\mathcal H].
\label{eq:R_Hcore_expanded}
\end{align}

Using the mismatch \(\mathcal M\) defined in
Eq.~\eqref{eq:Mmod_review}, Eqs.~\eqref{eq:L_Hcore_expanded} and
\eqref{eq:R_Hcore_expanded}, together with the vanishing diagonal of a
commutator, imply
\begin{equation}
\sum_i
m_{\nu_i}^2\,\mathcal M_{\alpha i}=0.
\label{eq:weighted_neutrino_sumrule}
\end{equation}
Unitarity also gives
\begin{equation}
\sum_i\mathcal M_{\alpha i}=0.
\end{equation}
For a nondegenerate neutrino spectrum, every row is therefore proportional
to the vector \(\boldsymbol{\mathcal N}\) in
Eq.~\eqref{eq:N_review}.

The columns obey an analogous constraint. Multiplying
Eq.~\eqref{eq:U_master} by \(U_\nu^\dagger\) from the left and by
\(U_\nu\) from the right gives
\begin{equation}
m_\nu U_\nu-U_\nu^\dagger m_\nu
=
i\eps\,U_\nu^\dagger H U_\nu .
\end{equation}
For \(\eps\neq0\), its diagonal, compared with that of
Eq.~\eqref{eq:U_master}, gives
\begin{equation}
(U_\nu^\dagger H U_\nu)_{ii}=H_{ii}.
\end{equation}
At \(\eps=0\), the same equality follows from \(U_\nu=S_\nu\).
Using
\begin{equation}
H=V_L^\dagger\hat m_\ell V_L,
\qquad
U_\nu^\dagger H U_\nu
=
V_R^{P\dagger}\hat m_\ell V_R^P,
\end{equation}
we obtain
\begin{equation}
\sum_\alpha
\hat m_\alpha\,\mathcal M_{\alpha i}=0.
\label{eq:weighted_charged_sumrule}
\end{equation}
Together with
\begin{equation}
\sum_\alpha\mathcal M_{\alpha i}=0,
\end{equation}
and the physical nondegeneracy of the charged-lepton masses, every column
is proportional to \(\boldsymbol{\mathcal E}\) in
Eq.~\eqref{eq:E_review}.

The row and column constraints prove the outer-product form
Eq.~\eqref{eq:M_outer_product_review}, showing that the modulus mismatch
has rank at most one.

The next spectral moment relates \(\Lambda\) to the scalar obstruction.
Equation~\eqref{eq:M_outer_product_review} gives
\begin{equation}
(R^2-L^2)_{\alpha\alpha}
=
\sum_i m_{\nu_i}^4\mathcal M_{\alpha i}
=
\Lambda\mathcal E_\alpha\mathcal D_\nu ,
\label{eq:Lambda_mass4}
\end{equation}
where \(\mathcal D_\nu\) is defined in Eq.~\eqref{eq:Dnu_review}.

From Eqs.~\eqref{eq:L_Hcore_expanded} and
\eqref{eq:R_Hcore_expanded},
\begin{equation}
R-L=i\eps[\hat m_\ell,\mathcal H],
\qquad
R+L=2\mathcal H^2+\frac{\eps^2}{2}\hat m_\ell^2 .
\end{equation}
Using
\begin{equation}
R^2-L^2
=
\frac12\{R-L,R+L\},
\end{equation}
the contribution proportional to \(\hat m_\ell^2\) vanishes on the
diagonal, and
\begin{equation}
(R^2-L^2)_{\alpha\alpha}
=
i\eps
\left\{
[\hat m_\ell,\mathcal H],
\mathcal H^2
\right\}_{\alpha\alpha}.
\end{equation}
For cyclic \((\alpha,\beta,\gamma)\), the three-generation identity
\begin{equation}
i\left\{
[\hat m_\ell,\mathcal H],
\mathcal H^2
\right\}_{\alpha\alpha}
=
2(\hat m_\beta-\hat m_\gamma)
\Im\!\left(
\mathcal H_{\alpha\beta}
\mathcal H_{\beta\gamma}
\mathcal H_{\gamma\alpha}
\right),
\end{equation}
then gives
\begin{equation}
(R^2-L^2)_{\alpha\alpha}
=
2\eps\mathcal E_\alpha
\mathcal I_{\mathcal H}.
\label{eq:IH_mass4}
\end{equation}
In the last step, the terms in \((\mathcal H^2)_{\beta\alpha}\) involving
diagonal entries of \(\mathcal H\) give real products and only the term passing
through the third flavor closes the CP-odd cycle.

Comparing Eqs.~\eqref{eq:Lambda_mass4} and
\eqref{eq:IH_mass4} yields
\begin{equation}
\Lambda\mathcal D_\nu
=
2\eps\mathcal I_{\mathcal H}.
\label{eq:Lambda_IH}
\end{equation}
Hence, for nondegenerate neutrino masses and \(\eps\neq0\),
\begin{equation}
\Lambda=0
\quad\Longleftrightarrow\quad
\mathcal I_{\mathcal H}=0.
\end{equation}
Together with Eq.~\eqref{eq:IH_equivalence_review}, this explains why
modulus alignment is already sufficient for full compatibility on the
physical parity image.
At \(\eps=0\), the modulus mismatch vanishes identically under parity and
therefore carries no CP information. The invariant
\(\mathcal I_{\mathcal H}\) remains the appropriate condition.

\subsection{Local equivalence of the Jarlskog condition}

Compatibility already implies \(J_R=-J_L\) through
Eq.~\eqref{eq:IH_equivalence_review}. We now determine when the converse
is locally true, so that the Jarlskog condition may be used by itself in
the analytic expansion.

For any \(3\times3\) Hermitian matrix
\begin{equation}
K
=
W\,\diag(k_1,k_2,k_3)\,W^\dagger ,
\end{equation}
the cubic identity
\begin{equation}
\Im\!\left(
K_{e\mu}K_{\mu\tau}K_{\tau e}
\right)
=
J(W)
(k_1-k_2)(k_2-k_3)(k_3-k_1)
\label{eq:cubic_J_identity}
\end{equation}
extracts the Jarlskog invariant of its diagonalizing matrix
\cite{Jarlskog:1985ht,Jarlskog:1985cw}.

Applied to \(L\) and \(R\), it gives
\begin{equation}
J_L\mathcal D_\nu
=
\Im\!\left(
L_{e\mu}L_{\mu\tau}L_{\tau e}
\right),
\quad
J_R\mathcal D_\nu
=
\Im\!\left(
R_{e\mu}R_{\mu\tau}R_{\tau e}
\right).
\label{eq:JLJR_cubic}
\end{equation}
For \(\alpha\neq\beta\),
\begin{align}
& L_{\alpha\beta}
=
(\mathcal H^2)_{\alpha\beta}
-
\frac{i\eps}{2}
(\hat m_\alpha-\hat m_\beta)
\mathcal H_{\alpha\beta},
\\
& R_{\alpha\beta}
=
(\mathcal H^2)_{\alpha\beta}
+
\frac{i\eps}{2}(\hat m_\alpha-\hat m_\beta)
\mathcal H_{\alpha\beta}.
\label{eq:LR_offdiag_compact}
\end{align}
The odd powers of \(\eps\) therefore cancel in \(J_R+J_L\). In three
generations the surviving CP-odd expression must be proportional to
\(\mathcal I_{\mathcal H}\). Direct cyclic expansion gives
Eq.~\eqref{eq:Jsum_IH_review}, with \(\Xi\) defined in
Eq.~\eqref{eq:Xi_review}.
The zeroth-order coefficient follows by applying
Eq.~\eqref{eq:cubic_J_identity} to \(\mathcal H\) and
\(\mathcal H^2\). The quadratic terms combine into the second term in
Eq.~\eqref{eq:Xi_review}.

Combining Eqs.~\eqref{eq:Jsum_IH_review} and \eqref{eq:Lambda_IH},
without dividing by \(\mathcal I_{\mathcal H}\), gives
\begin{equation}
\Lambda\Xi=\eps(J_R+J_L).
\label{eq:Lambda_Xi_review}
\end{equation}
The modulus mismatch and the Jarlskog orientation are therefore two
observable projections of the same scalar obstruction.

At \(\eps=0\), the eigenvalues of \(\mathcal H\) are the signed neutrino
masses, and Eq.~\eqref{eq:Xi_seed_review} follows directly.
Away from an exact signed degeneracy, \(\Xi\neq0\) in a neighbourhood of
either CP-conserving solution. The Jarlskog condition is then locally
equivalent to the full compatibility requirement, as stated in
Eq.~\eqref{eq:J_equivalence_review}.

\bibliographystyle{ws-mpla}
\interlinepenalty=10000
\bibliography{refs}

\end{document}